\documentstyle[eqsecnum,floats,aps]{revtex}

\newcommand{\tr}{\mbox{tr}}

\newcommand{\ee}{\end{equation}}
\newcommand{\be}{\begin{equation}}

\newcommand{\vep}{\varepsilon}

\newcommand{\re}[1]{(\ref{#1})}

\newcommand{\bCL}{{\cal L}\hspace{-1.51ex}{\cal L}}
\newcommand{\bCC}{{\cal C}\hspace{-1.25ex}{\cal C}}
\newcommand{\bCF}{{\cal F}\hspace{-1.8ex}{\cal F}}

\renewcommand{\d}{{\rm d}}

\newcommand{\Z}{{\rm Z \hspace{-0.75ex} Z}}
\newcommand{\R}{{\rm I \hspace{-0.52ex} R}}

\newcommand{\eins}{1\hspace{-0.56ex}{\rm I}}

\newcommand{\CE}{{\cal E}}

\newcommand{\CF}{{\cal F}}

\newcommand{\bomega}{\mbox{\boldmath $\omega$}}

\newcommand{\balpha}{\mbox{\boldmath $\alpha$}}

\newcommand{\sphere}{\kappa}

\newcommand{\isomo}{\cong}
\newcommand{\x}{\rm x}
\newcommand{\overlap}{H}

\newcommand{\bPhi}{{\Phi}}
\newcommand{\bF}{{\bf F}}
\newcommand{\bD}{{\bf D}}
\newcommand{\bA}{{\bf A}}

\newcommand{\bOmega}{{\bf \Omega}}
\newcommand{\sOmega}{\bar{\bOmega}}
\newcommand{\dsOmega}{{}^{\pm}\!\bar{\bOmega}}

\newcommand{\rF}{{\rm F}}
\newcommand{\rD}{{\rm D}}
\newcommand{\rA}{{\rm A}}
\newcommand{\rS}{{\rm S}}

\newcommand{\dPhi}{{}^{\pm}\!\bPhi}

\newcommand{\dA}{{}^{\pm}\!\bA}

\newcommand{\sPhi}{\bar{\bPhi}}
\newcommand{\sA}{\bar{\bA}}
\newcommand{\sF}{\bar{\bF}}
\newcommand{\sD}{\bar{\bD}}

\newcommand{\dsPhi}{{}^{\pm}\!\bar{\bPhi}}
\newcommand{\dsA}{{}^{\pm}\!\bar{\bA}}
\newcommand{\dsF}{{}^{\pm}\!\bar{\bF}}

\newcommand{\aPhi}{\bPhi_{(P)}}

\newcommand{\aA}{\bA_{(P)}}

\newcommand{\saPhi}{\sPhi_{(P)}}
\newcommand{\saF}{\sF_{(P)}}
\newcommand{\saA}{\sA_{(P)}}

\newcommand{\daPhi}{{}^{\pm}\!\aPhi}

\newcommand{\daA}{{}^{\pm}\!\aA}

\newcommand{\dsaPhi}{{}^{\pm}\!\saPhi}
\newcommand{\dsaA}{{}^{\pm}\!\saA}
\newcommand{\dsaF}{{}^{\pm}\!\saF}

\newcommand{\tB}{\mbox{\textsf{\textbf{H}}}}
\newcommand{\tA}{\mbox{\textsf{\textbf{B}}}}
\newcommand{\tG}{{\bf \Lambda}}
\newcommand{\ttG}{\mbox{\textsf{\textbf{U}}}}

\newcommand{\mtB}{\tB_{(P)}}

\newcommand{\dtA}{{}^{\pm}\!\tA}
\newcommand{\dmtA}{{}^{\pm}\!\tA_{(P)}}

\newcommand{\tcB}{\mbox{\textsf{H}}}
\newcommand{\tcA}{\mbox{\textsf{B}}}
\newcommand{\tcG}{\Lambda}

\newcommand{\tdB}{\tilde{\tcB}}
\newcommand{\tdA}{\tilde{\tcA}}
\newcommand{\tdG}{\tilde{\tcG}}

\newcommand{\gm}{\gamma_{\mu}}
\newcommand{\gn}{\gamma_{\nu}}
\newcommand{\gmn}{\gamma_{\mu\nu}}
\newcommand{\gd}{\gamma_{d}}
\newcommand{\gi}{\gamma_{i}}
\newcommand{\gij}{\gamma_{ij}}
\newcommand{\gik}{\gamma_{ik}}

\newcommand{\hx}{\hat{x}}
\newcommand{\hxm}{\hat{x}_{\mu}}
\newcommand{\hxn}{\hat{x}_{\nu}}
\newcommand{\hxd}{\hat{x}_{d}}
\newcommand{\hxi}{\hat{x}_{i}}
\newcommand{\hxj}{\hat{x}_{j}}
\newcommand{\hxk}{\hat{x}_{k}}

\newcommand{\g}{{\rm g}}
\newcommand{\G}{{\rm T}}

\newcommand{\sG}{\bar{\G}}

\newcommand{\aG}{\G_{(P)}}
\newcommand{\saG}{\sG_{(P)}}

\newcommand{\gpm}{\g_{\pm}}

\newcommand{\ah}{\varepsilon_{d+1}}

\newcommand{\degree}{{\cal D}}
\newcommand{\degnorm}{{D}}

\newcommand{\winding}{{\cal W}}
\newcommand{\windnorm}{{W}}

\begin{document}

\title{'t~Hooft tensors as Kalb--Ramond fields of generalised monopoles
in all odd dimensions: $d=3$ and $d=5$}

\author{ {\large
D. H. Tchrakian}\footnote{e-mail: tigran@thphys.may.ie}$^{a)b)}$
and {\large F. Zimmerschied}\footnote{e-mail: zimmers@thphys.may.ie}$^{a)}$}

\address{{$^{a)}$\small Department of Mathematical Physics,
National University of Ireland Maynooth, Maynooth, Ireland}}
\address{{\small and}}
\address{{$^{b)}$\small School of Theoretical Physics -- DIAS, 10 Burlington 
Road, Dublin 4, Ireland }}

\maketitle

\begin{abstract}

The Kalb--Ramond monopole, as discussed by Nepomechie, is identical with the (singular) Dirac monopole 
in $d=3$ dimensions. The latter can be described by the (regular) 't~Hooft--Polyakov
monopole, via the 't~Hooft tensor construction. This construction is extended to arbitrary
odd dimensions by performing the $d=5$ case explicitly, exploiting the (regular) `monopoles'
of generalised Georgi--Glashow models and identifying their 't~Hooft tensors as the Kalb--Ramond fields. 
The relevant `magnetic charges' are expressed as topological invariants.
\end{abstract}
\medskip
\medskip

\medskip


\medskip
\medskip

\section{Introduction}

In {\it three} Euclidean space dimensions, a Dirac~\cite{D} magnetic monopole is a singular 
static solution of $U(1)$ Maxwell electromagnetism with nonvanishing magnetic
flux over a sphere surrounding the monopole. Dirac's original description of 
such an object involved a string--like singularity, extending from the
location of the monopole to infinity. An important step in the description of the Dirac 
monopole which avoids the ``Dirac string'' involves Wu--Yang~\cite{WY} fields on overlapping
coordinate patches (which in mathematical terms correspond to a nontrivial 
$U(1)$ fibre bundle over $S^2$), but the magnetic field strength still 
exhibits a singularity at the location of the monopole.

'tHooft~\cite{tH} showed that the soliton solution of the $d=3$ Georgi--Glashow (GG)
model, which is known as 't~Hooft--Polyakov monopole, 
can be interpreted as a regular realisation of the Dirac magnetic monopole,
identifying the unbroken $U(1)$ subgroup of the soliton with Maxwell 
magnetism and the topological soliton charge with magnetic charge. This
is done using the 't Hooft tensor which is identified with the magnetic
$U(1)$ curvature of the 't~Hooft--Polyakov~\cite{tH,P} monopole outside the
core. In regular gauge, the 'tHooft tensor supports a description of the
topology and hence the magnetic charge in terms of the Higgs field only~\cite{AFG}, 
whereas in singular gauge, the 'tHooft tensor restores the Wu--Yang
description of the Dirac monopole.

Dirac's monopole construction in Maxwell theory (which is a theory
of antisymmetric rank two tensor field strengths, i.e.\ two--forms)
was generalised by Nepomechie \cite{N} to Kalb--Ramond (KR) theory \cite{KR} 
which itself generalises Maxwell theory to higher rank antisymmetric tensor
field strengths. The resulting KR monopoles in higher dimensions are 
descibed by singular KR field strengths, and a Wu--Yang type construction 
allows the introduction of KR potentials for the monopoles without string 
singularities. As KR theory coincides with ordinary Maxwell theory
in three spatial dimensions, Nepomechie's construction~\cite{N} really generalises
Dirac's original work to higher dimensions. In fact the KR generalisation of the
Dirac monopole to arbitrary dimensions was performed earlier in the context of lattice
field theory by Savit~\cite{S}, Orland~\cite{O} and Pearson~\cite{Pe}. The motivation of
these authors was to construct a dilute 'instanton' gas in arbitrary dimensionsl
Abelian models, following previous work of Polyakov~\cite{Pol}. The construction of
dilute gases will naturally arise as a possible application of the present work, and
will be alluded to in our Discussions.

It is our intention in the present work, to supply regular solitonic realisations of
Nepomechie's KR monopoles and to generalise the construction of the 't~Hooft tensor
to higher dimensions. This is done concretely for dimension $d=5$, in a sufficiently
general framework which points clearly to the systematic generalisation to {\it all
odd dimensions}.

To realise our objective, there are two main ingredients needed. The first is a natural
generalisation of the GG model in higher dimensions, which supports solitonic solutions
stabilised by a topological charge. The second is the construction of a
generalisation of the 't~Hooft electromagnetic tensor, which describes the Abelian field
strength to be identified with the higher dimensional KR field. 
The first of these, namely the generalised GG (gGG) models, are readily obtained by
subjecting members of the hierarchy of Yang-Mills (YM) models in $4p$ dimensions~\cite{T}
to dimensional descent. Examples of the resulting Higgs models in various dimensions can be
found in Refs.~\cite{T&OC}. The second, namely the definition of 't~Hooft tensors in all
odd dimensions, is entirely new. We have done this concretely
in $d=5$, and shown that such a 't~Hooft tensor can be implemented in {\it odd} dimensions only.
The construction relies on the proper definition of the
C-P densites that present the lower bounds on the relevant 
gGG models. These C-P densities are obtained by subjecting the
original Chern--Pontryagin (C-P) densities in $4p$ dimensions, to dimensional descent~\cite{T&OMS}.

In Section {\bf II}, we consider the gauge and Higgs fields, along with the gauge group, its
representation and the Higgs multiplets, that we need if we are to satisfy the requisite
topological properties necessary for the desired constructions. This includes also a description
of the {\it regular} and {\it singular} (Dirac) gauges. Section {\bf III} is divided into two
subsections. In the first we discusses the candidates for the generalised GG models that can be
employed, while in the second, the corresponding C-P charge densities descended from
the C-P densities in higher dimensions are discussed. Section {\bf IV}, in which we present the main
results, is divided into three subsections. In the first, we prove that the C-P charge equals the
Higgs field winding number in regular gauge in all dimensions, and that the C-P charge can be
interpreted as magnetic KR~\cite{N} charge in odd dimensions.
In the second Subsection, we use the Dirac gauge to construct Wu--Yang type
KR potentials for the C-S forms and interpret the solitonic solutions to the odd dimensonal
models as regular realisations of KR monopoles. This motivates the nomenclature `gGG monopoles' for
these regular solutions. In the third Subsection we construct the generalised 't~Hooft 
tensors  which we identify with the KR field strengths, and show in which sense these 't~Hooft tensors 
contain the (odd dimensional) results of the two previous Subsections.
Section {\bf V} is devoted to a discussion of our results. In Appendix {\bf A}
we give a brief description of KR fields and Nepomechie's KR monopole constructions in all
dimensions, that is relevant to the present work. In Appendix {\bf B} we list the action/energy
functionals of the four simplest gGG models, as well as the relevant C-P densities and their
Chern--Simons (C-S) forms.

\section{Gauge groups and topology}
\label{secmodels}

Our primary considerations in this work are the topological properties of gauged Higgs systems,
and their relation to KR monopoles in odd dimensions. To this end
we set up the topological
framework by selecting the required gauge groups, their representations, as well as the
representations of the Higgs fields. Irrespective of the detailed dynamics, which is discussed in
the next Section, we can impose the
finite action/energy conditions which are expected to lead to topologically nontrivial configurations,
and which yield the asymptotic fields. We will discuss these asymptotic fields both in
the {\it regular} and the {\it singular} (Dirac) gauges. 

In $d$ Euclidean dimensions, we consider a $d$ vector multiplet Higgs field $\phi^{\sigma}$
which we write in isovector representation
\begin{equation}
\bPhi = \phi^{\sigma}\ah\gamma_{\sigma}
\label{higgs}
\end{equation}
where $\{\gamma_1,\ldots,\gamma_d\}$ are the Euclidean gamma matrices in $d$
dimensions, and $\ah$ is an ``antihermitean factor''. In even dimensions,
$d=2M$, there exists a chiral matrix 
$\gamma_{2M+1}=\gamma_{d+1}=i\gamma_1\cdots\gamma_{2M}$ which is used as 
antihermitean factor, $\varepsilon_{2M+1}=\gamma_{2M+1}$, whereas no chiral 
matrix exists in odd dimensions $d=2M+1$, and the imaginary unit $i$ is
used instead, $\varepsilon_{2M+2}=i$, hence 
\begin{equation}
\ah := \left\{ \begin{array}{l}
\gamma_{d+1} \qquad \mbox{($d$ even)} \\
i \qquad\quad\;\; \mbox{($d$ odd)} 
\end{array} \right. .
\label{ah}
\end{equation}
The Higgs fields under consideration are gauged with $SO(d)$ gauge potentials $\bA$ 
taking values in the $so(d)$ algebra with antihermitean
generators $\gmn=-\frac14[\gm,\gn]$,
\begin{equation}
\bA = A_{\mu}^{[\rho\sigma]}\gamma_{\rho\sigma}\d x^{\mu}.
\label{gauge}
\end{equation}
Boldface letters here denote forms, in components 
$\bA=\rA_{\mu}\d x^{\mu}$. The corresponding field strength is
$\bF=\d\bA+\bA\wedge\bA=\frac12{\rm F}_{\mu\nu}\d x^{\mu}\wedge \d x^{\nu}$.

Both Higgs and gauge fields can be assumed to be regular on $\R^d$, a
property wich is not destroyed by regular {\em gauge transformations} 
given by $SO(d)$--valued functions $\g=\exp\left(R^{\mu\nu}\gmn\right)$ 
on $\R^d$ with
\begin{eqnarray}
\bPhi & \mapsto & {}^{\g}\bPhi :=
\g\bPhi\g^{-1}  \\
\bA & \mapsto & {}^{\g}\bA := 
\g\bA\g^{-1}+\g\d\g^{-1}\: .
\label{gaugetrans}
\end{eqnarray}
Physical quantities, in particular the energy functional defining a concrete 
theory, have to be invariant under regular gauge transformations.

The existence of an energy functional yields further constraints
on the fields considered, because it involves an integral which has to 
converge for a given field configuration. Therefore, finite energy 
configurations are characterised by a particular asymptotic behaviour of the 
fields $(\bPhi,\bA)$. Denoting the asymptotic fields (i.e.\ the 
leading terms of an asymptotic $\frac1r$ expansion of $\bPhi$ and $\bA$)
by $\sPhi$ and $\sA$, appropriate finite energy conditions for a large
class of models, including those discussed in Section {\bf III}, read
\begin{eqnarray}
\sPhi^2 & = & -\eins \label{CSOa}\\
\sD\sPhi &:=&\d\sPhi+[\sA,\sPhi] = 0.
\label{CS0}
\end{eqnarray}
Conditions \re{CSOa} and \re{CS0} anticipate the general features of the generalised GG models
to be introduced in Subsection {\bf III A} below.

If the fields $(\bPhi,\bA)$ are regular on $\R^d$, then the asymptotic fields
labeled by overbars, $(\sPhi,\sA)$, are defined on 
$S^{d-1}$, and (\ref{CS0}) can be solved for asymptotic gauge potential,
\begin{equation}
\d\sPhi+[\sA,\sPhi]=0\quad\Rightarrow\quad
\sA = -\frac{1}{4}[\sPhi,\d\sPhi],
\label{cs5}
\end{equation}
hence the asymptotic configurations in regular gauge are determined by
the Higgs field alone which at infinity yields a mapping
\begin{equation}
\sPhi: S^{d-1}\rightarrow S^{d-1}\: .
\label{top1}
\end{equation}
Therefore, any finite energy configuration in regular gauge can be 
topologically classified in terms of the homotopy group 
\begin{equation}
\Pi_{d-1}(S^{d-1})\isomo\Z
\label{top2}
\end{equation}
using the (integer) {\em winding number} of the (asymptotic) Higgs field 
\begin{equation}
\winding^{(d)}\{\bPhi\} = \frac{1}{\windnorm_d}\int_{S^{d-1}}\tr\big[\ah
\sPhi\underbrace{\d\sPhi\wedge\ldots\wedge\d\sPhi}_{d-1}\big]
\label{defwinding}
\end{equation}
as topological invariant, where $\windnorm_d$ is a normalisation constant. 

The simplest example with these topological properties are the radially 
symmetric configurations which in regular gauge are given by
\begin{equation}
\aPhi = \ah  h(r) \gm\hxm, \qquad 
\aA = \frac{1+f(r)}{r} \gmn\hxn\d x^{\mu} \: ,
\label{Ansatz}
\end{equation}
with $x_{\mu}=r\hx_{\mu}$ and $\hx_{\mu}^2 =1$.
Requiring finite energy, the profile functions 
$h$ and $f$ have to satisfy $h(r\rightarrow\infty)=1$,
$f(r\rightarrow\infty)=0$, whereas regularity at the origin means 
$h(0)=0$, $f(0)=-1$. 
>From a topological point of view, the particular shape 
of $h(r)$, $f(r)$ is not important since it does not affect the 
asymptotic radially symmetric fields
\begin{equation}
\saPhi = \ah\gm\hxm, \qquad
\saA = \frac{1}{r}\gmn\hxn \d x^{\mu}
\label{a2}
\end{equation}
as well as the asymptotic field strength given by
\begin{equation}
\saF = - \frac{1}{2r^2}\left(\gmn+\hx_{[\mu}\gamma_{\nu]\lambda}\hx_{\lambda}
\right)\d x^{\mu}\wedge\d x^{\nu}. 
\label{a2a}
\end{equation}
Obviously, interpreted as the mapping  
\begin{equation}
\saPhi: S^{d-1}\rightarrow S^{d-1}
\label{a2b}
\end{equation}
the asymptotic Higgs field of any radially symmetric configuration
(\ref{Ansatz}) has winding number 1, $\winding^{(d)}\{\saPhi\}=1$. Higher winding number requires 
Ans\"atze with more involved, e.g.\ axial, symmetry properties.

Besides the regular gauge transformations discussed so far, there are also
singular gauge transformations which are defined only on some subset of
$\R^d$, the corresponding gauge transformed fields also being defined only
on that subset and singular on the (generally point-- or string--like)
complement. Those singular gauge transformations are of particular interest
because they allow the transformation of the fields to the {\em Dirac gauge} 
in which the Higgs field always points in the isospace $d$--direction. In particular,
the asymptotic Higgs field of finite energy configurations trivialises 
to $\pm\ah\gd$ in the Dirac gauge which allows the characterisation of two distinct Dirac
gauges, {\em positive} and {\em negative}. Using this characterisation, (positive or negative) Dirac 
gauge transformations are regular on 
\begin{equation}
\R^d_{\pm}:=\R^d\setminus\{(0,\ldots,0,\mp x)|x\ge 0\}
\label{aaa3}
\end{equation}
but singular on either the positive or the negative $d$--axis. Therefore, 
the fields $(\dPhi,\dA)$ in positive or negative Dirac gauge are also 
singular on the negative or positive $d$--axis, respectively, commonly known as ``Dirac string''.

To describe a finite energy configuration $(\bPhi,\bA)$ on $\R^d$
in Dirac gauge, one always needs $({}^+\!\bPhi,{}^+\!\bA)$ in 
positive Dirac gauge, defined on $\R^d_+$, as well as 
$({}^-\!\bPhi,{}^-\!\bA)$ in negative Dirac gauge, defined on  
$\R^d_-$. Both $({}^+\!\bPhi,{}^+\!\bA)$ and $({}^-\!\bPhi,{}^-\!\bA)$
have to be gauge equivalent to $(\bPhi,\bA)$. It follows that on the 
overlap of the positive and negative Dirac gauge definition ranges, 
$\R^d_0:=\R^d_+\cap\R^d_-$, there exists a {\em transition gauge 
transformation} $\G$ with 
\begin{equation}
{}^{\G}({}^-\!\bPhi,{}^-\!\bA) = ({}^+\!\bPhi,{}^+\!\bA)\: .
\label{ttrans}
\end{equation}
Asymptotically, the trivialisation of the Higgs field in Dirac gauge
no longer allows one to express the asymptotic gauge field in terms of
the asymptotic Higgs field since (\ref{cs5}) requires regularity of the
fields on $\R^d$. Instead, the finite energy condition (\ref{CS0}) 
forces the breaking of the gauge symmetry of the asymptotic gauge field according to
\begin{equation}
\d\dsPhi+[\dsA,\dsPhi]=0\quad\Rightarrow\quad
[\dsA,\gd]=0\quad\Rightarrow\quad \dsA = {}^{\pm}\!\bar{A}^{[ij]}_{\mu}
\gij\d x^{\mu}\: 
\label{dd1}
\end{equation}
hence $\dsA$ takes values in the $so(d-1)$ subalgebra of $so(d)$ and is defined on 
$S^{d-1}\setminus\{0,\ldots,0,\mp 1\}$ which is the $d-1$ dimensional sphere ``at infinity'', 
excluding the south or north pole, respectively. To describe 
the asymptotic gauge fields on $S^{d-1}$, it is sufficient to consider $\dsA$ on the
upper or lower half spheres $S^{d-1}_{\pm}$, respectively, which overlap on the
{\em equator} $S^{d-2}=S^{d-1}_+\cap S^{d-1}_-$.

In regular gauges, finite energy configurations could be topologically
classified in terms of the asymptotic Higgs field winding number $\winding^{(d)}\{\bPhi\}$. 
In the Dirac gauge, the asymptotic Higgs fields
$\dsPhi=\pm\ah\gd$ do not carry any topological information,
but the fact that one needs both positve and negative Dirac gauge to describe 
a single finite energy configuration now yields the topological characterisation.
This can be be expressed in terms of the asymptotic transition gauge 
transformation (\ref{ttrans}) which reverses the sign 
of the asymptotic Higgs field, hence
\begin{equation}
\sG(-\ah\gd)\sG^{-1} = \ah\gd\quad\Rightarrow\quad \{\sG,\gd\} = 0.
\label{dpm2}
\end{equation}
$\sG$ takes values in the subset $\overlap(d)\subset SO(d)$ defined by 
(\ref{dpm2}), $\overlap(d)\isomo S^{d-2}$,
and transforms ${}^-\!\bA$ to ${}^+\!\bA$ on the overlap of
their definition ranges which is the equator $S^{d-2}$. This means that $\sG$ is topologically equivalent 
to a mapping
\begin{equation}
\sG: S^{d-2}\rightarrow S^{d-2},
\label{top3}
\end{equation}
which enables the classification of a finite energy configuration in the
Dirac gauge in terms of the homotopy group 
\begin{equation}
\Pi_{d-2}(S^{d-2})\isomo \Z,
\label{top4}
\end{equation}
expressed by the degree of the (asymptotic) transition gauge transformation
\begin{equation}
\degree^{(d)}\{\G\} = \frac{1}{\degnorm_d}\int_{S^{d-2}} \tr\big[\vep_d
\underbrace{\left(\sG\d\sG^{-1}\right)\wedge\ldots\wedge\left(\sG\d\sG^{-1}\right)}_{d-2}\big]
\label{defdegree}
\end{equation}
where $\degnorm_d$ is a normalisation constant.

Considering the example of the radially symmetric field configuration discussed
above, the singular gauge transformations
\begin{equation}
\gpm = \frac{1}{\sqrt{2(1\pm\hxd)}}\left\{(1\pm\hxd)\eins\pm\gd\hxi\gi\right\},
\label{a3}
\end{equation}
which are well defined on $\R^d_{\pm}$, respectively, transform 
$(\aPhi,\aA)$ in eq.\ (\ref{Ansatz}) to the positive or negative Dirac gauges,
$(\daPhi,\daA)={}^{\gpm}(\aPhi,\aA)$. The asymptotic fields are
\begin{eqnarray}
{}^{\gpm}\left(\saPhi\right)
& := & \dsaPhi = \pm \ah\gd 
\label{a4a} \\
{}^{\gpm}\left(\saA\right) & := & \dsaA = \frac{1}{r}
\frac{1}{1\pm\hxd}\gij\hxj\d x^i, 
\label{a4b}
\end{eqnarray}
with gauge field strength,
\begin{equation}
\dsaF = -\frac{1}{2r^2}\left(\gij+\frac{1}{1\pm\hxd}
\hat{x}_{[i}\gamma_{j]k}\hxk\right)\d x^i\wedge \d x^j
\pm \frac{1}{r^2}\gik\hxk\d x^i\wedge \d x^d.
\label{a5}
\end{equation}
The explicit expressions (\ref{a4b},\ref{a5}) show 
that the gauge group of the asymptotic hedgehog configuration in Dirac gauge breaks 
down as $SO(d)\rightarrow SO(d-1)$, whereas the asymptotic 
Higgs field (\ref{a4a}) trivialises.

The transition gauge transformation $\aG$, 
${}^{\aG}({}^-\!\aPhi,{}^-\!\aA) = ({}^+\!\aPhi,{}^+\!\aA)$, is in this case given by
\begin{equation}
\aG = \g_+\g_-^{-1} =  \frac{1}{\sqrt{1-\hxd^2}}\gd\hxi\gi \: .
\label{a7}
\end{equation}
Restricting $\saG=\aG$ to the equator $\hxd=0$ yields
\begin{equation}
\saG\big|_{\hxd=0}=\gd\hxi\gi \: ,
\label{a6}
\end{equation}
which is a mapping $S^{d-2}\rightarrow S^{d-2}$ of degree $1$, $\degree^{(d)}\{\saG\}=1$.

\section{Models and topological charges}

In the previous Section we selected the gauge group to be $SO(d)$ and the representation of the
Higgs field to be the $d$ component vector. This choice was made on topological criteria,
including the possibility of having a Dirac gauge. Here, we further require that models like
these must also support solitonic solutions which means that the action/energy is bounded from below by a
topological charge. In any given dimension $d$, there are in principle an infinite number of
such models, out of which is is reasonable to select the simplest one. These are all derived
from members of the $4p$ dimensional YM hierarchy~\cite{T} with the gauge field in one of the
two chiral representations of $SO_{\pm}(4p)$, whose action density is given by
\be
\label{1}
\bCL^{(4p)}=\tr \underbrace{\left(\bCF\wedge\ldots\wedge\bCF\right)}_{2p}\wedge
*\underbrace{\left(\bCF\wedge\ldots\wedge\bCF\right)}_{2p}
\ee
where $\bCF$ is the $so(4p)$ valued $2$--form gauge field strength (curvature), and $*$
denotes the Hodge dual. As in $4$ dimensions, the action density \re{1} is bounded from 
below by the $2p$-th C-P density $\bCC_{2p}$
\be
\label{2}
\int \bCL^{(4p)}\: \ge \: \int \bCC_{2p} \: .
\ee
It is known that when \re{2} is saturated, the resulting self--duality equations have both
spherically~\cite{sph} and axially~\cite{ax} symmetric solutions.

The derived $d$-dimensional gauged Higgs models result from the dimensional descent of the inequality
\re{2} over some $(4p-d)$-dimensional compact coset space $K^{4p-d}$
\be
\label{3}
\int_{\R^d \times K^{4p-d}} \bCL^{(4p)}\: \: \ge \: \:
\int_{\R^d \times K^{4p-d}} \bCC_{2p} \: .
\ee
After performing the (compact) integration over the $K^{4p-d}$ coordinates, we are left with
the inequality for the $d$-dimensional action/energy density of the residual gauged Higgs model,
bounded from below by residual C-P density. We will discuss these two quantities
in a little more detail in the following two Subsections. Before proceeding however, we make
two remarks.

Firstly, the choice of the compact coset space $K^{4p-d}$ is not important for our purposes
since we are not concerned with the gauge coupling constant explicitly, so we will have in mind
the simplest variant $K^{4p-d}=S^{4p-d}$ when discussing the symmetry breaking that occurs in
the dimensional descent, and refer to the corresonding residual models as generalised Georgi--Glashow (gGG)
models.

Secondly, we will restrict to the simplest of all possible residual system in any dimension $d$.
It is clear from \re{3} that the descent to $d$ dimensions can start from any dimension $4p>d$.
The simplest systems will result, obviously, when $4p$ is the smallest number that is larger
than $d$. We shall refer to these as the {\it minimal} gGG models.
In our considerations below, we will always restrict ourselves to these choices.
Familiar examples of such residual models are the three dimensional GG model (in the
Prasad-Somerfield limit) and the Abelian Higgs model, both descended from $SO_{\pm}(4)=SU(2)$  
$p=1$ (i.e. usual) YM\footnote{If by contrast the descent to $d=3$ is started from the $p=2$ YM,
another variant of the GG model is obtained~\cite{KOT}. Its asymptotic properties are identical to 
those of the usual GG model and hence it yields nothing new in the present context.
We therefore exclude all such models from consideration here.}.

\subsection{The generalised GG models and finite energy conditions}

Having explained the general procedure used in the derivation of residual Higgs models above, we now
discuss some properties of minimal gGG systems in $d=2,3,4,5$ which are used in the subsequent Section.
They arise respectively, from the dimensional reduction of the $p=1$,
$SO_{\pm}(4)=SU(2)$ ($d=2,3$), and the $p=2$, $SO_{\pm}(8)$ ($d=4,5$), members of
the YM hierarchy, and are given explicitly in the Appendix {\bf B}. We have denoted the 
energy/action density of the residual models by $\CE^{(p,d)}$ and will refer to them as
energy densities henceforth, since in
the familiar 3 dimensional case this coincides with the definition of an energy density.
We note that the $d=2$ model $\CE^{(1,2)}$ thus obtained is nothing but the usual 
Abelian Higgs model, whereas $\CE^{(1,3)}$ equals the GG model in the Prasad--Somerfield limit, but we
do not make a distinction on this acount because this limit makes no difference to our considerations
below, the latter being sensitive only to the asymptotic {\it values} rather than the detailed decays
of the fields.

An important general feature of the
energy densities of these residual models is, that the curvature $\bF$, the covariant derivative
of the Higgs field $\bD\bPhi$ and the 'square root' of the Higgs selfinteraction potential $\rS =
-(\bPhi^2 +\eins)$
all must decay asymptotically at the same rate to satisfy
{\it finite energy conditions}. The reason for this is easily explained. Denoting the curvatures
in $4p$ dimensions that appear in \re{1} with indices $M=\mu,m$, with $\mu$ labeling the coordinates of
$\x\in\R_d$ and $m$ those of ${\rm y}\in S^{4p-d}$, as a result of the dimensional reduction we have,
for $d$ {\em odd} and {\em even} respectively,
\be
\label{indices}
\CF_{MN} =\left\{ 
\begin{array}{l}
\CF_{\mu \nu}=\rF_{\mu \nu} \otimes Y({\rm y}) \\
\CF_{\mu m}=\rD_{\mu}\bPhi \otimes Y_m({\rm y}) \\
\CF_{mn}=\rS\otimes Y_{mn}({\rm y})
\end{array}\right., \quad
\CF_{MN} =\left\{ 
\begin{array}{l}
\CF_{\mu \nu}=\rF^{(+)}_{\mu \nu} \otimes Y^{(+)}({\rm y}) +
\rF^{(-)}_{\mu \nu} \otimes Y^{(-)}({\rm y}) \\
\CF_{\mu m}=\rD_{\mu}\varphi \otimes Y^{(+)}_m({\rm y})-
\rD_{\mu}\varphi^{\dagger} \otimes Y^{(-)}_m({\rm y}) \\
\CF_{mn}=s_+ \otimes Y^{(+)}_{mn}({\rm y}) + s_- \otimes Y^{(-)}_{mn}({\rm y})
\end{array}\right.
\ee
where $s_+ =\varphi \varphi^{\dagger} -\eins$, $s_- =\varphi^{\dagger} \varphi -\eins$, and the
$Y$'s, are ${\rm y}\in S^{4p-d}$ dependent  tensor--spinor bases whose details do not concern us here
(see for example Refs.~\cite{T&OC} for details). In the even dimensional cases, the Higgs
field $\bPhi$, and the gauge potential $\rA_{\mu}$ are composed of the corresponding fields $\varphi$,
$\varphi^{\dagger}$ and $\rA_{\mu}^{(\pm)}$ (whose curvatures are $\rF_{\mu \nu}^{(\pm)}$) as follows
\be
\label{comps}
\rA_{\mu}=\left[
\begin{array}{cc}
\rA_{\mu}^{(+)} & 0 \\
0 & \rA_{\mu}^{(-)}
\end{array}
\right]\: , \qquad
\bPhi=\left[
\begin{array}{cc}
0 & \varphi \\
-\varphi^{\dagger} & 0
\end{array}
\right]\: .
\ee
What is interesting here is that the substitution of \re{indices} into \re{1}, which yields
the residual energy density in $d$ dimensions, results in a sum of terms each of which consists
of $2p$ factors of all possible types of components listed in \re{indices}. It follows that each of the
fields $\rF_{\mu \nu}$, $\rD_{\mu}\bPhi$ and $(\bPhi^2 +\eins)$ must have the same asymptotic decay 
rate if the energy is to be finite. This means that $\rD_{\mu}\bPhi$ and $\rS=-(\bPhi^2 +\eins)$
can be neglected in asymptotic expansions which justifies the conditions \re{CSOa} and \re{CS0}
defining the asymptotic fields $(\sPhi,\sA)$.

Another important property of the models under consideration is that they support nontrivial, stable
finite energy solutions to which we will refer to as solitons in accord with our nomenclature
$\CE^{(p,d)}$ as energy density. In particular, the radially symmetric 
Ansatz (\ref{Ansatz}) minimises the energy functionals, i.e.\ the profile functions $f(r)$ and $h(r)$ 
can be chosen such that they solve the Euler--Lagrange equations of the radial subsystem (which in general 
requires numerical integration techniques). The corresponding radially symmetric solution is called 
`hedgehog'.

\subsection{Topological charges of generalised monopoles}

Under dimensional reduction, the left hand side of \re{3} yields the residual subsystems which are the
candidates for gGG models. The right hand side, which is the dimensionally reduced C-P density, presents 
the lower bounds on the gGG energy densities,
\begin{equation}
\CE^{(p,d)} \ge \varrho^{(p,d)}\: .
\label{gg1}
\end{equation}
The volume integral over this C-P density is called C-P charge $q^{(p,d)}$,
\begin{equation}
q^{(p,d)} = \int_{\R^d}\varrho^{(p,d)} d^dx \: .
\label{gg2}
\end{equation} 
The normalisation of $\varrho^{(p,d)}$ and $\CE^{(p,d)}$ in \re{gg1} is chosen such that the hedgehog
\re{Ansatz} has unit C-P charge. $q^{(p,d)}$ is commonly also called topological charge of the gGG
model $\CE^{(p,d)}$ since it is closely related to the topological properties of finite energy 
configurations discussed in Section {\bf II}. This is shown in Subsection {\bf IV A}, making use of the
most important property of the C-P densities $\varrho^{(p,d)}$, namely that they are total divergences, 
\begin{equation}
\varrho^{(p,d)}=\partial_{\lambda}\Omega^{(p,d)}_{\lambda}.
\label{gg3}
\end{equation}
This was shown in numerous cases in Refs.~\cite{T&OMS}, both for even and odd values of the 
residual dimensions $d$, which we do not exhibit here, except for the four examples discussed in 
Appendix {\bf B}. $\Omega^{(p,d)}_{\lambda}$ is the residual C-S density and  
$\bOmega^{(p,d)}=\Omega^{(p,d)}_{\lambda}(*\d x^{\lambda})$ the residual C-S form which we refer to as 
the {\em C-S form} pertaining to the model $\CE^{(p,d)}$, in the sense that it allows us to write the 
C-P charge (\ref{gg2}) as a surface integral over the boundary of $\R^d$,
\begin{equation}
q^{(p,d)} = \lim_{r\rightarrow\infty}\int_{S^{d-1}(r)}\hx_{\lambda} \Omega^{(p,d)}_{\lambda} dS =
\lim_{r\rightarrow\infty}\int_{S^{d-1}(r)} \bOmega^{(p,d)}.
\label{gg4}
\end{equation}
We point out that these {\em C-S forms}, are {\em gauge invariant in odd,  and gauge
variant in even, dimensions}~\cite{T&OMS}. This is clearly seen from
\re{CS:p=1,d=2}--\re{CS:p=2,d=5} of Appendix
{\bf B}. This property of these C-S forms will be very important in our subsequent considerations.

For use in the next Section, we now introduce the asymptotic expressions of C-S forms
$\bOmega^{(p,d)}$ appearing in \re{gg4}. We denote them, again with an overbar, as
\begin{equation}
\sOmega^{(d)}:=\bOmega^{(p,d)}\Big|_{(\sPhi,\sA)}\: ,
\label{gg5}
\end{equation}
where $(\sPhi,\sA)$ are the asymptotic fields defined in \re{CSOa} and \re{CS0}. We have labeled
$\sOmega^{(d)}$ with the residual dimension $d$ and not with the label $p$ that specifies the model,
since it does not depend on the latter. This is because of
the general structure of the asymptotic C-S forms which 
is discussed below. Accordingly the C-P charge, evaluated by the surface integral \re{gg4}, also is 
independent of $p$ and we express it as
\begin{equation}
q^{(d)} = \int_{S^{d-1}(r)} \sOmega^{(d)}.
\label{surface}
\end{equation}
The asymptotic C-S forms $\sOmega^{(d)}$ inherit the important
property of the $\bOmega^{(p,d)}$, namely that they are gauge invariant for odd $d$ and gauge variant for
even $d$. Accordingly we treat the two cases separately.

In $2M+1$ {\em (odd) dimensions}, the asymptotic C-S forms for the two examples $d=3$ and $d=5$
which will be needed below, can be readily extracted from the general C-S forms
\re{CS:p=1,d=3} and \re{CS:p=2,d=5} given in Appendix {\bf B}, using the finite energy requirements. They are
\begin{eqnarray}
\sOmega^{(3)} & = & \frac{1}{c_3}
\tr\left[\sPhi\sF\right] \label{gg6a} \\
\sOmega^{(5)} & = & \frac{1}{c_5}
\tr\left[\sPhi\sF\wedge\sF\right] \label{gg6b}.
\end{eqnarray}
In general, $\bOmega^{(p,2M+1)}$ consists of products of $\bF$, $\bD\bPhi$ and $\rS$. As a consequence of
the finite energy conditions, its asymptotic form $\sOmega^{(2M+1)}$ consists of $M$ factors $\bF$ and one
Higgs field. By virtue of the identity
\begin{equation}
\tr[\bD\bPhi\wedge\underbrace{\bF\wedge\ldots\wedge\bF}_{M}] = \d\,
\tr[\bPhi\underbrace{\bF\wedge\ldots\wedge\bF}_{M}]
\label{gg7}
\end{equation}
then, the asymptotic C-S form in odd dimensions takes the form
\begin{equation}
\sOmega^{(2M+1)} = \frac{1}{c_{2M+1}}\tr[\sPhi\underbrace{\sF\wedge\ldots\wedge\sF}_{M}]
\label{gg8}
\end{equation}
where $c_{2M+1}$ is a normalisation constant. From eq.\ (\ref{gg8}), it is obvious that 
$\sOmega^{(2M+1)}$ is gauge invariant.

In $2M$ {\em (even) dimensions}, following similar arguments as before now applied to \re{CS:p=1,d=2} and
\re{CS:p=2,d=4}, yield the asymptotic C-S forms for the $d=2$ and $d=4$ dimensional models,
\begin{eqnarray}
\sOmega^{(2)} & = & \frac{1}{c_2}
\tr\left[\gamma_3\sA\right] =:  \frac{1}{c_{2}}\bomega^{(1)}[\sA,\sF]  \label{gg9a} \\
\sOmega^{(4)} & = & \frac{1}{c_4}
\tr\left[\gamma_5\left(\sF\wedge\sA-\frac13\sA\wedge\sA\wedge\sA\right)
\right] =: \frac{1}{c_{4}}\bomega^{(2)}[\sA,\sF]  \label{gg9b}.
\end{eqnarray}
In \re{gg9a} and \re{gg9b}, we have introduced a new symbol $\bomega^{(M)}$ used in Ref.~\cite{Zumino},
for $M=1$ and $M=2$, because is will be useful in the work of subsequent sections for general $M$.
In general, $\bOmega^{(p,2M)}$ consists of products of $\bF$, $\bD\bPhi$, $\rS$, and of $\bA$.
As a consequence of the finite energy conditions, its asymptotic form $\sOmega^{(2M)}$ consists only of
the Higgs independent terms, and equals the trace of the chiral $SO(2M)$ matrix $\gamma_{2M+1}$ times the
products of $\bA$ and $\bF$ that appear in the trace of the C-S form of the (chiral) $SO_{\pm}(2M)$
Yang-Mills fields without a Higgs field.
Using $[\bA,\gamma_{2M+1}]=0$ one can show \cite{Zumino} that this is an exact form,
\begin{equation}
\tr[\gamma_{2M+1}\underbrace{\bF\wedge\ldots\wedge\bF}_{M}] = \d\bomega^{(M)}[\bA,\bF]\: .
\label{gg10}
\end{equation}
Unlike the corresponding expression \re{gg8} which can be expressed for all $d=2M+1$, the corresponding
expressions in even dimensions have a more complicated dependence on $d=2M$, abbreviated by
$\bomega^{(M)}$. These are easy to find
explicitly in any given case and for $d=6$ and $d=8$ can be found in the first item in Refs.\cite{T&OMS}.
One can show 
that it is this term which contributes the leading asymptotic behaviour of the C-S form,
hence the asymptotic C-S from in even dimensions can be written as
\begin{equation}
\sOmega^{(2M)} = \frac{1}{c_{2M}}\bomega^{(M)}[\sA,\sF]\: .
\label{gg11}
\end{equation}

Eqs.~\re{gg8} and \re{gg11} show that the asymptotic C-S form really depends on the
dimension only and not on the particular model characterised by $p$, the member of the YM hierarchy
the underlying model was derived from.

We emphasise that the asymptotic C-S forms  in odd dimensions always
involve one Higgs field in addition to an antisymmetric curvature $M$-form, while in even dimensions
the Higgs field is absent and is replaced by the chiral $SO(d)$ matrix.

\section{Kalb-Ramond fields and 't~Hooft tensors}

This Section is divided in three Subsections. In the first we evaluate the C-P charge in
the regular gauge identifying it with the winding number of the Higgs field $\winding^{(d)}\{\bPhi\}$,
and finding that in odd
dimensions this gives rise to the description of magnetic KR~\cite{N} charge. The second subsection is
devoted to the corresponding considerations in the Dirac gauge. There, we have shown that the C-P
charge in even dimensions cannot be evaluated as a surface integral in the Dirac gauge, while that is
possible in odd dimensions. It was also shown that the C-P charge results from a surface integral
over Wu-Yang type KR potentials, and that it can be computed as the degree $\degree^{(d)}\{\G\}$
of the transition gauge
transformation between positive and negative Dirac gauges. Finally in the third Subsection, we have
given the prescription to construct 't~Hooft tensors for all odd dimensional monopoles as a natural
extension of the content of the previous two Subsections.

\subsection{Regular gauge}
\label{CPreg}

The C-P charge can be evaluated directly by simply inserting a field configuration
in regular gauge into the surface integral (\ref{surface}). For the hedgehog field configurations
$(\aPhi,\aA)$, (\ref{Ansatz}), the asymptotics (\ref{a2}) yield
\begin{equation}
\sOmega^{(d)}_{(P)} := \bOmega^{(p,d)}\Big|_{(\saPhi,\saA)} =
\frac{1}{\sphere_{d-1}}\frac{\hxm}{r^{d-1}} (*\d x^{\mu})\qquad \Rightarrow \qquad
\int_{S^{-1}} \sOmega^{(d)}_{(P)} = 1\: ,
\label{KRnorm}
\end{equation}
in which $\sphere_{d-1}$ denotes surface of $S^{d-1}$.
This direct evaluation does not, however, demonstrate the topological nature of the C-P charge,
i.e. its relation to the topological invariants of a finite energy configuration discussed in 
Section {\bf II}.

In regular gauge, a relation between the C-P charge $q^{(d)}$ and the
winding number of the Higgs field $\winding^{(d)}\{\bPhi\}$ is easy to find. Using (\ref{cs5})
to rewrite the asymptotic gauge fields in terms of the asymptotic Higgs fields,
\begin{equation}
\sA = -\frac12 \sPhi\d\sPhi, \qquad \sF = -\frac14\d\sPhi\wedge\d\sPhi
\label{CPreg1}
\end{equation}
and inserting (\ref{CPreg1}) into \re{gg6a}, \re{gg6b}, \re{gg9a}, \re{gg9b} we find 
\begin{eqnarray}
\sOmega^{(2)} & = & -\frac{1}{2c_2}\,\tr\left[\gamma_3\sPhi\d\sPhi\right]
\label{CPreg2a} \\
\sOmega^{(3)} & = & -\frac{1}{4c_3}\,\tr\left[\sPhi\d\sPhi\wedge\d\sPhi\right]
\label{CPreg2b} \\
\sOmega^{(4)} & = & \hspace{0.7ex}
\frac{1}{6c_4}\,\tr\left[\gamma_5\sPhi\d\sPhi\wedge\d\sPhi\wedge\d\sPhi\right]
\label{CPreg2c} \\
\sOmega^{(5)} & = & \hspace{0.7ex}
\frac{1}{16c_5}\,\tr\left[\sPhi\d\sPhi\wedge\d\sPhi\wedge\d\sPhi\wedge\d\sPhi\right]\: .
\label{CPreg2d}
\end{eqnarray}
By virtue of \re{CPreg2a}--\re{CPreg2d}, the C-P charge $q^{(d)}$ can be identified with the
winding number of the Higgs field defined by (\ref{defwinding}),
\begin{equation}
q^{(d)} = \winding^{(d)}\{ \bPhi \},
\label{CPreg3}
\end{equation}
in both even and odd dimensions. This justifies our description of C-P charges as topological charges
in the previous Section.
We stress that (\ref{CPreg1}) is valid in regular gauge only,
therefore, eq.\ (\ref{CPreg3}) can not be used to express the C-P charge in Dirac gauge by performing 
a singular gauge transformation, according to $\winding^{(d)}\{\dPhi\} = 0 \neq  q^{(d)}$.

Because a factor of one Higgs field appears in the odd dimensional asymptotic C-S forms
\re{gg8}, it follows that these are {\em closed} forms, namely that 
\begin{equation}
\d\sOmega^{(2M+1)} = \frac{1}{c_{2M+1}}\tr[\sD\sPhi\underbrace{\sF\wedge\ldots\wedge\sF}_{M}]
= 0\: .
\label{CPreg4}
\end{equation}
We should emphasise that this result follows from the presence in \re{CPreg4} of $\sD\sPhi$ which
is vanishing asymptotically according to \re{CS0}.

In even dimensions however, the expression corresponding to \re{CPreg4} is
\begin{equation}
\d\sOmega^{(2M)} = \frac{1}{c_{2M}}\d\bomega^{(M)}[\sA,\sF]=
\frac{1}{c_{2M}}\tr[\gamma_{2M+1}\underbrace{\sF\wedge\ldots\wedge\sF}_{M}] \neq 0 \: ,
\label{CPreg5}
\end{equation}
which is {\em nonvanishing}, i.e. $\sOmega^{(2M)}$ {\em is not a closed form}.

Being a {\em closed form}, the asymptotic C-S form in $2M+1$ dimensions can be identified with
a {\em Kalb--Ramond field strength} (cf.\ Appendix {\bf A}) on $S^{2M}$
which supports magnetic flux equal to the C-P charge, $Q_m =q^{(2M+1)}$.

This allows us to interpret the C-P charge $q^{(d)}$ solitons of the gGG model
with energy density $\CE^{(p,2M+1)}$ 
({\it e.g.} the hedgehog \re{Ansatz} with $q^{(d)}=1$) as the KR monopole \re{KR13}
with magnetic charge $Q_m = q^{(d)}$. In other words, we can describe the singular KR monopole field
as the $2M+1$ dimensional (regular) soliton of the appropriate gGG model.

\subsection{Dirac gauge}

The situation in Dirac gauge is different from that in regular gauge in so far as it is not possible
to evaluate the C-P charge in {\em even dimensions} using the surface integral
(\ref{surface}), a fact already emphasised in Ref.~\cite{MT}. 
For example, inserting the hedgehog (\ref{Ansatz}) into the asymptotic C-S
form $\sOmega^{(2M)}$ yields
\begin{equation}
\int_{S^{2M-1}}\dsOmega^{(2M)}_{(P)} = 0,
\label{CPdir2}
\end{equation}
where $\dsOmega^{(2M)}_{(P)}$ is the asymptotic C-S form in the positive or negative Dirac
gauge for the hedgehog field configuration given by \re{a4a} and \re{a4b}.
This is because of the gauge variance of $\sOmega^{(2M)}_{(P)}$ under the large gauge transformations
(\ref{a3}) which take the regular
hedgehog to positive or negative Dirac gauge. We will demonstrate this in detail in $2M=2$ and $2M=4$
and will then give the general case.

In $2M=2$ dimensions where
\be
\label{z1}
\dsOmega^{(2)}_{(P)}= \frac{1}{c_2}\tr\left[\gamma_3\dsaA\right]
\ee
we find
\begin{equation}
\dsOmega^{(2)}_{(P)} =  
\sOmega^{(2)}_{(P)} +\frac{1}{c_2}
\tr\left[\gamma_3\gpm\d\gpm^{-1}\right] =0\: .
\label{CPdir4}
\end{equation}
It turns out that the two terms in \re{CPdir4} simply cancel out.

In $2M=4$ dimensions, where
\be
\label{z2}
\dsOmega^{(4)}_{(P)}= \frac{1}{c_4}
\tr\left[\gamma_5\left(\dsaF\wedge\dsaA-\frac13\dsaA\wedge\dsaA\wedge\dsaA\right) \right]
\ee
we find
\begin{equation}
\dsOmega^{(4)}_{(P)} =  
\sOmega^{(4)}_{(P)} +\frac{1}{c_4}\left\{\d\, \tr\left[\gamma_5\saA\wedge\d\gpm^{-1}\right]
-\frac13 \tr\left[\gamma_5\left(\gpm\d\gpm^{-1}\right)\wedge\left(\gpm\d\gpm^{-1}\right)
\wedge\left(\gpm\d\gpm^{-1}\right)\right]\right\}=0
\label{CPdir3a}
\end{equation}
where again the sum of terms in \re{CPdir3a} cancel out.

Finally in general where, with the notation of \re{gg11}
\be
\label{z3}
\dsOmega^{(2M)}_{(P)}= \frac{1}{c_{2M}}\bomega^{(M)}[\dsaA,\dsaF]
\ee
the C-S form in Dirac gauge can be rewritten as
\begin{equation}
\dsOmega^{(4)}_{(P)} =  
\sOmega^{(4)}_{(P)} +\frac{1}{c_{2M}}\left\{
\d \balpha^{(M)}[\sA,\g] +
\bomega^{(M)}[\gpm\d\gpm^{-1},0]\right\}.
\label{CPdir5}
\end{equation}
Here, we have used the transformation properties of $\bomega^{(M)}$ in the notation of
Ref.~\cite{Zumino}. Note that the general expression \re{CPdir5} is exactly of the same form as the
special case \re{CPdir3a} for $M=2$, while the $M=1$ case \re{CPdir4} is of a simpler form. Note also
that in \re{CPdir5}, the exact form $\d \balpha^{(M)}[\sA,\g]$ does not contribute in the surface
integral \re{CPdir2}, in which case the other two terms cancel out.

In {\em odd dimensions} on the other hand, the surface integral of $\dsOmega^{(2M+1)}$ over $S^{2M}$
does yield the correct value for the C-P charge \cite{MT}, like in the regular gauge. In this case
the asymptotic C-S form of a finite energy configuration 
$(\dPhi,\dA)$ in Dirac gauge, can be given explicitly for the general case. As a consequence of the
asymptotic form of the Higgs field in this gauge, $\dsPhi=\pm i\gamma_{2M+1}$, and \re{gg8},
\begin{equation}
\dsOmega^{(2M+1)} = \pm\frac{i}{c_{2M+1}}
\tr\big[\gamma_{2M+1}\underbrace{\dsF\wedge\ldots\wedge\dsF}_{M}\big]\: .
\label{CPdir10}
\end{equation}
This quantity is defined on the sphere $S^{2M}\setminus \{(0,\ldots,0,\mp 1)\}$ excluding the
south or north pole, respectively. As a consequence of the manifest gauge invariance of \re{CPdir10},
\be
\sOmega^{(2M+1)} = \dsOmega^{(2M+1)} .
\label{CPdir1}
\ee

Another important property of the C-S forms $\sOmega^{(2M+1)}$, besides their gauge invariance, 
is that they are closed forms. This was used in Subsection {\bf IV A} to interpret them as
of KR field strengths on $S^{2M}$. Moreover in this case, namely in the Dirac gauge, the expression
\re{CPdir10} makes it possible to express the KR fields as {\em curls} of KR potentials
on the simply connected regions $S^{2M}_{\pm}$. Following Nepomechie~\cite{N}, we  denote these KR
potentials by $\dtA^{(2M+1)}$, 
\begin{equation}
\sOmega^{(2M+1)} = \d\dtA^{(2M+1)}
\label{KRpot}
\end{equation}
in the notation of Appendix {\bf B}. We will construct these Wu--Yang type
KR potentials explicitly in three and five dimensions first and will then present the general case. 

In $2M+1=3$ dimensions, the gauge group of $\dsF$ breaks down to $SO(2)$ by virtue of 
$[\gamma_3,\dsF]=0$, as discussed in Section {\bf II}. We then have 
\begin{equation}
\dsOmega^{(3)} = \pm\frac{i}{c_3}\tr\left[\gamma_3\dsF\right] = \pm\frac{i}{c_3}\d\, 
\tr\left[\gamma_3\dsA\right],
\label{zz1}
\end{equation}
which yields
\begin{equation}
\dtA^{(3)} = \pm\frac{i}{c_3} \tr\left[\gamma_3\dsA\right]. 
\label{zz2}
\end{equation}
In $2M+1=5$ dimensions, $\dsF$ has $SO(4)$ gauge symmetry by virtue of $[\gamma_5,\dsF]=0$, from which it
follows that
\begin{equation}
\dsOmega^{(5)} = \pm\frac{i}{c_5}\tr\left[\gamma_5\dsF\wedge\dsF\right] = \pm\frac{i}{c_5}\d\, 
\tr\left[\gamma_{5}
\left(\dsF\wedge\dsA-\frac13\dsA\wedge\dsA\wedge\dsA\right)
\right]\: ,
\label{zz3}
\end{equation}
thus
\begin{equation}
\dtA^{(5)} = \pm\frac{i}{c_5} \tr\left[\gamma_{5}
\left(\dsF\wedge\dsA-\frac13\dsA\wedge\dsA\wedge\dsA\right)
\right]. 
\label{zz4}
\end{equation}
We now note that the forms on the right hand sides of \re{zz1} and \re{zz3} coincide with the forms
$\bomega^{(1)}$ and $\bomega^{(2)}$ in the corresponding even dimensional ($d=2,4$) cases \re{gg9a} and
\re{gg9b}. In general $2M+1$ dimensions, $\dsF$ is a $SO(2M)$ gauge field strength according to \re{dd1}.
This enables us to write the KR potentials of the asymptotic
C-S forms in general $2M+1$ dimensions with the help of $\bomega^{(M)}$ (which was
used in (\ref{gg11}) in the context of even dimensional C-S forms) as follows
\begin{equation}
\dtA^{(2M+1)}=\pm\frac{i}{c_{2M+1}} \bomega^{(M)}\left[\dsA,\dsF\right].
\label{CPdir12}
\end{equation}

The KR potentials $\dtA^{(2M+1)}$ can be used to rewrite the C-P charge surface integral
corresponding to \re{surface}, over $S^{2M}$, as an integral over the equator $S^{2M-1}$.
First splitting the integral \re{surface} into two
integrals over the upper and lower half spheres $S^{2M}_{\pm}$, respectively, one can use the gauge
invariance of the odd dimensional C-S forms, \re{CPdir1}, to evaluate the C-P charge $q^{(2M+1)}$
without integrating over the singularities in the Dirac gauge. Finally, using the KR potentials
\re{KRpot} and applying Stoke's theorem on $\partial S^{2M}_+ = - \partial S^{2M}_- = S^{2M-1}$
yields
\begin{equation}
q^{(2M+1)}= \int_{S^{2M}_+}{}^+\!\sOmega^{(2M+1)} + \int_{S^{2M}_-}{}^-\!\sOmega^{(2M+1)}=
\int_{S^{2M-1}}\left({}^+\!\tA^{(2M+1)} - {}^-\!\tA^{(2M+1)}\right)
\label{CPdir16}
\end{equation}
According to the discussion in Appendix {\bf A 2}, the integral
over the {\em transition form} 
\begin{equation}
\ttG^{(2M+1)}={}^+\!\tA^{(2M+1)}-{}^-\!\tA^{(2M+1)}
\label{CPdirtrans}
\end{equation}
in \re{CPdir16} determines the KR magnetic charge of the the Wu--Yang type KR potentials $\dtA^{(2M+1)}$
which enables us again to identify the C-P charge $q^{(2M+1)}$ with magnetic KR charge $Q_m$.

In the context of the gGG models, the transition form $\ttG^{(2M+1)}$ is a composite field
which can be written explicitly in terms of the asymptotic gauge potentials $\dsA$ and the 
asymptotic transition gauge transformation $\sG$ which takes the negative to the positive Dirac gauge, 
${}^{\sG}({}^-\!\sPhi,{}^-\!\sA) = ({}^+\!\sPhi,{}^+\!\sA)$.  Again, we first present this result
in the three and five dimensional case, giving the general odd dimensional result afterwards.

In {\em three dimensions}, using $\{\sG,\gamma_3\}=0$, \re{dpm2}, we find
\be
{}^+\!\tA^{(3)} = \frac{i}{c_3} \tr\left[\gamma_3{}^{\sG}{}\!\left(^-\!\sA\right)\right]
= {}^-\!\tA^{(3)} + \frac{i}{c_3}\tr\left[\gamma_3\sG\d\sG^{-1}\right]\: ,
\label{zzz1}
\ee
yielding $\ttG^{(3)} = \frac{i}{c_3}\tr\left[\gamma_3\sG\d\sG^{-1}\right]$. Substituting this into the
equatorial integral \re{CPdir16} leads to
\be
\label{CPdir19a}
q^{(3)}  =  \frac{i}{c_{3}}\int_{S^{1}}\tr\left[\gamma_3\sG\d\sG^{-1}\right]\: .
\ee

In {\em five dimensions} we find in the same way, using $\{\gamma_5,\sG\}=0$,
\be
\ttG^{(5)} = \frac{i}{c_5}\left\{\d\, \tr\left[\gamma_5^-\!\sA\wedge\d\sG^{-1}\right]
-\frac13 \tr\left[\gamma_5\left(\sG\d\sG^{-1}\right)\wedge\left(\sG\d\sG^{-1}\right)
\wedge\left(\sG\d\sG^{-1}\right)\right]
\right\}\: .
\label{zzz4}
\ee
Substituting this into the equatorial (closed surface) integral \re{CPdir16}, by Stokes' theorem
the first term in \re{zzz4} does not contribute and the rest yields
\be
\label{CPdir19b}
q^{(5)} = -\frac{i}{3c_{3}}\int_{S^{3}}\tr\left[\gamma_5\left(\sG\d\sG^{-1}\right)
\wedge\left(\sG\d\sG^{-1}\right)\wedge\left(\sG\d\sG^{-1}\right)\right]\: .
\ee

The five dimensional example already shows the structure of the general case. Using the
same notations $\balpha^{(M)}$ and $\bomega^{(M)}$ introduced in \re{CPdir5} 
but taking care of the anticommutator$\{\gamma_{2M+1},\sG\}=0$,
\begin{equation}
\ttG^{(2M+1)}=\frac{i}{c_{2M+1}}\left\{\d\balpha^{(M)}\left[{}^-\!\sA,\sG\right] +
\bomega^{(M)}\left[\sG\d\sG^{-1},0\right]\right\}.
\label{zzz5}
\end{equation}
Substituting this into the equatorial integral \re{CPdir16}, in which the term
$\d\balpha^{(M)}\left[{}^-\!\sA,\sG\right]$ does not contribute,
we can write the C-P charge in the general case as
\begin{equation}
q^{(2M+1)} = \frac{i}{c_{2M+1}}\int_{S^{2M-1}}\bomega^{(M)}\left[\sG\d\sG^{-1},0\right]\: .
\label{CPdir18}
\end{equation}

Finally, we see that the C-P charges \re{CPdir19a} and \re{CPdir19b} equal, respectively, the degrees of the
transition gauge transformations \re{defdegree}, in three and five dimensions. In the general case with
$q^{(2M+1)}$ given by \re{CPdir18}, we have
\begin{equation}
q^{(2M+1)} = \degree^{(2M+1)}\{\G\}\: .
\label{CPdir20}
\end{equation}

\subsection{Construction of the 't~Hooft tensor}

In this Subsection, we give the prescription for constructing {\em 't~Hooft tensors}~\cite{tH,AFG}
pertaining to the gGG monopoles, in all odd dimensions. This encapsulates the
results of both previous Subsections {\bf IV A} and {\bf IV B} simultaneously.
In three dimensions, the 't~Hooft tensor is identified with 
the residual Maxwell field of the Dirac monopole. In general odd dimensions, the Dirac
monopole generalises to KR monopoles described in terms of KR fields (which coincide with Maxwell fields
in $d=3$ dimensions). The construction of 'tHooft tensors in higher dimensions then follows naturally. 

We first repeat the construction in three dimensions, using the notation developed 
above and pointing out the main steps as a guideline for the five dimensional case presented
subsequently. This construction can be systematically extended to all odd dimensions.

In {\em three dimensions}, the asymptotic C-S form \re{gg6a} can be written as
\begin{equation}
\sOmega^{(3)} = \frac{1}{c_3}
\tr\left[\sPhi\sF\right] = \frac{1}{c_3}\tr\left[\sPhi\left(\d\sA+\sA\wedge\sA\right)\right]
\label{th1}
\end{equation}
The plan now is to add an asymptotically vanishing, gauge invariant, expression which cancels the
second (highest order in $\sA$'s) term in \re{th1}, such that the result consists of
an exact form plus a term depending on the Higgs fields only. To be concrete, one can check the identity
\be
0 = \tr\left[\sPhi\sD\sPhi\wedge\sD\sPhi\right]
= \tr\left[\sPhi\left(\d\sPhi\wedge\d\sPhi
+ 4\sA\wedge\sA\right) 
- 4\d\sPhi\wedge\sA\right]
\label{th2}
\ee 
which is asymptotically zero by virtue of $\sD\sPhi=0$. Multiplying the identity \re{th2} by the 
normalisation factor $-(4c_3)^{-1}$, and adding this to the asymptotic C-S form \re{th1}, we have the desired
definition of the usual 't~Hooft tensor
\begin{eqnarray}
\tB^{(3)} & := & \sOmega^{(3)} - \frac{1}{4c_3}\tr\left[\sPhi\sD\sPhi\wedge\sD\sPhi\right]
\label{th4aa} \\
& = &   \d\tA^{(3)} -\frac{1}{4c_3}\tr\left[\sPhi\d\sPhi\wedge\d\sPhi\right]
\label{th4}
\end{eqnarray}
with
\be
\label{th4a}
\tA^{(3)} = \frac{1}{c_3}\tr\left[\sPhi\sA\right]\: .
\ee
In the definition equation \re{th4}, we have used the symbol $\tB^{(3)}$ for the 't~Hooft tensor, which
is also the symbol for the KR field strength introduced in Appendix {\bf A}. This is because $\sOmega^{(3)}$
on the left hand side of \re{th1} describes both these quantities, and the 't~Hooft tensor can be identified
with the KR field strength. The C-P charge now equals the surface integral of this KR field strength 
$\tB^{(3)}$, which by virue of \re{KR7}, defines magnetic KR charge,
\be
q^{(3)} = \int_{S^2} \sOmega^{(3)} = \int_{S^2} \tB^{(3)} := Q_m \: .
\label{th5}
\ee

In the {\em regular} gauge, where the surface integral is over the entire (closed) $S^2$, the term
$\d\tA^{(3)}$ in \re{th4} does not contribute. It follows that the only contribution comes from the 
second term
\be
q^{(3)} = \int_{S^2} \tB^{(3)} = -\frac{1}{4c_3}\int_{S^2}\tr\left[\sPhi\d\sPhi\wedge\d\sPhi\right]
=\winding^{(3)}[\bPhi]\: .
\label{th6}
\ee
This reproduces the result \re{CPreg3} of Subsection {\bf IV A} in three dimensions.

In the {\em Dirac} gauge, the surface integral is carried out over the sphere with a hole around the
negative or positive (south or north) Dirac string singularity, and the Higgs field is a constant.
As a consequence of the constancy of the Higgs
field, the term $\tr\left[\sPhi\d\sPhi\wedge\d\sPhi\right]$ in \re{th4} vanishes so that
the only contribution to the surface integral comes from the residue of the singularity in
the first term. This is the (closed $S^1$) line integral of
\be
\tA^{(3)}\big|_{(\dsPhi,\dsA)}=
\pm\frac{i}{c_3}\tr\left[\gamma_3\dsA\right] = \dtA^{(3)}\: ,
\label{th7}
\ee
in the notation of \re{zz2}. Here $\dtA^{(3)}$ are the KR potentials of the KR field strength
$\tB^{(3)}= \d\dtA^{(3)}$ on $S^2_{\pm}$. This reproduces the results of Subsection {\bf IV B}.

In {\em five dimensions}, the asymptotic C-S form \re{gg6b} can be written as
\begin{eqnarray}
\sOmega^{(5)} & = & 
\frac{1}{c_{(5)}}\tr\left(\sPhi\sF\wedge\sF\right)  \nonumber \\
& = &   
\frac{1}{c_{(5)}}
\tr\left[\sPhi\left(\d\sA\wedge\d\sA + \d\sA\wedge\sA\wedge\sA + \sA\wedge\sA\wedge\d\sA +
\sA\wedge\sA\wedge\sA\wedge\sA\right)\right].
\label{th8}
\end{eqnarray}
As in the three dimensional case, the plan now is to add an asymptotically vanishing, gauge invariant,
expression which cancels the fourth (highest order in $\sA$'s) term in \re{th8}, such that the result
consists of an exact form plus a term depending on the Higgs fields only. Unlike in the three dimensional
case where there is only one such candidate, {\it c.f.} $\tr\left[\sPhi\sD\sPhi\wedge\sD\sPhi\right]$ in
eq.~\re{th2}, now there are two, namely 
$\tr\left[\sPhi\sD\sPhi\wedge\sD\sPhi\wedge\sD\sPhi\wedge\sD\sPhi\right]$ 
and $\tr\left[\sPhi\sF\wedge\sD\sPhi\wedge\sD\sPhi\right]$. It only remains to find the relative numerical
coefficients of these two terms, such that they result in the elimination of the said term.
To be concrete, one can check the identity
\begin{eqnarray}
0=\tr\left[\sPhi\left(\sD\sPhi\wedge\sD\sPhi-8\sF\right)\wedge
\sD\sPhi\wedge\sD\sPhi\right]
& = &  
\tr\left[\sPhi\d\sPhi\wedge\d\sPhi\wedge\d\sPhi\wedge\d\sPhi\right]
-16\,  \tr\left[\sPhi\sA\wedge\sA\wedge\sA\wedge\sA\right] 
\nonumber \\
&& {}  - 8 \, \d\,\tr\left[\sPhi\left(\d\sPhi\wedge\d\sPhi+\d\sPhi\wedge\sA\sPhi 
+ \mbox{$\frac13$}\sA\sPhi\wedge\sA\sPhi\right)\wedge
\sA\right]
\nonumber \\
&& 
{} - 8 \, \tr\left[\sPhi\left(\d\sA\wedge\sA\wedge\sA+
\sA\wedge\d\sA\wedge\sA+\sA\wedge\sA\wedge\d\sA\right)\right]
\nonumber \\
&& 
{} + 8 \, \tr\left[\d\sPhi\left(\d\sA\wedge\sA + \sA\wedge\d\sA +  \sA\wedge\sA\wedge\sA\right)\right].
\label{th9}
\end{eqnarray}
Multiplying the identity \re{th9} by the normalisation factor $(16c_3)^{-1}$, and adding this to the 
asymptotic C-S form \re{th8}, we have the definition of the 't~Hooft tensor in five dimensions,
\begin{eqnarray}
\tB^{(5)} & := & \sOmega^{(5)} + \frac{1}{16c_5} 
\tr\left[\sPhi\left(\sD\sPhi\wedge\sD\sPhi-8\sF\right)\sD\sPhi\wedge\sD\sPhi\right] 
\label{th10a} \\
& = & \d\,\tA^{(5)} + \frac{1}{16c_5} \tr\left[\sPhi\d\sPhi\wedge\d\sPhi\wedge\d\sPhi\wedge\d\sPhi\right]
\label{th10}
\end{eqnarray}
with
\be
\tA^{(5)}=\frac{1}{2c_{(5)}}\,\tr\left[\sPhi\left(
\d\sA\wedge\sA+\sA\wedge\d\sA + \sA\wedge\sA\wedge\sA-
\frac13\sA\wedge\sPhi\sA\sPhi\wedge\sA\right) +
\d\sPhi\wedge\left(\sPhi\d\sPhi+\sPhi\sA\sPhi\right)\wedge
\bA\right]
\label{th11}
\ee
Again, we have used the symbol $\tB^{(3)}$ in the definition equation \re{th4} which
is also the symbol for the KR field strength introduced in Appendix {\bf A}, defining the
KR field strength describing the monopoles of the gGG models as the
't~Hooft tensor \re{th10}. The C-P charge then is the surface integral of the KR field strength 
$\tB^{(5)}$ over $S^4$ in analogy to \re{th5},
\be
q^{(5)}=\int_{S^4} \sOmega^{(5)} = \int_{S^4} \tB^{(5)} = Q_m\: .
\label{th12}
\ee

In the {\em regular} gauge, where the surface integral is over the entire (closed) $S^4$, the term
$\d\tA^{(5)}$ in \re{th10} does not contribute. It follows that the only contribution comes from the 
second term
\be
q^{(5)} = \int_{S^4} \tB^{(5)} = \frac{1}{16c_5}\int_{S^4}
\tr\left[\sPhi\d\sPhi\wedge\d\sPhi\wedge\d\sPhi\wedge\d\sPhi\right]
=\winding^{(5)}[\bPhi]\: .
\label{th13}
\ee
This reproduces the result \re{CPreg3} of Subsection {\bf IV A} in five dimensions.

In the {\em Dirac} gauge, the surface integral is carried out over the sphere with a hole around the
negative or positive (southor north) Dirac string singularity, and the Higgs field is a constant.
As a consequence of the constancy of the Higgs
field, the term $\tr\left[\sPhi\d\sPhi\wedge\d\sPhi\wedge\d\sPhi\wedge\d\sPhi\right]$ in \re{th10}
vanishes so that the only contribution to the surface integral comes from the residue of the singularity in
the first term. This is the (closed $S^3$) integral of
\be
\tA^{(5)}\big|_{(\dsPhi,\dsA)}=
\pm\frac{i}{c_5}\tr\left[\gamma_{5}
\left(\dsF\wedge\dsA-\frac13\dsA\wedge\dsA\wedge\dsA\right)\right] = \dtA^{(5)}
\label{th14}
\ee
in the notation of \re{zz4}. $\dtA^{(5)}$ are the KR potentials of the KR field strength
$\tB^{(5)}=\d\dtA^{(5)}$ on $S^4_{\pm}$. This reproduces the result of Subsection {\bf IV B}.

It is now clear that the prescription for the construction of 
the 't~Hooft tensor in three and five dimensions
can be systematically extended to any odd dimension. The result can be stated formally, omitting the
expression corresponding to \re{th4aa} and \re{th10a} 
in the definitions of the 't~Hooft tensors, and restricting
to the result corresponding to \re{th4} and \re{th10}. It is 
\be
\label{tht1}
\tB^{(2M+1)} = \d \tA^{(2M+)} + \frac{1}{W_{2M+1}} \tr\big[\sPhi\underbrace{\d\sPhi\wedge\ldots
\wedge\d\sPhi}_{M}\big],
\ee
where $W_{2M+1}=(-4)^Mc_{2M+1}$ is the normalisation constant appearing in \re{defwinding}. 
In the Dirac gauge,
the expression corresponding to \re{th7} and \re{th14} is
\be
\label{tht2}
\tA^{(2M+1)}\big|_{(\dsPhi,\dsA)} = \pm\frac{i}{c_{2M+1}}\bomega^{(M)}[\dsA,\dsF] = 
\dtA^{(2M+1)}
\ee
which yields the KR potential \re{CPdir12} on $S^{2M}_{\pm}$ for the KR field strength $\tA^{(2M+1)}$ 
which we generally define as the 't~Hooft tensor of an odd dimensional gGG model.

\section{Summary and Discussion}

We have studied the topological properties of the solitons of the generalised
Georgi--Glashow (gGG) models, which we have 
described along with the corresponding reduced Chern--Pontryagin (C-P) densities and
Chern--Simons (C-S) forms in the {\em regular} and the the {\em Dirac} gauges, in all dimensions.
In common with the familiar three dimensional case, the topology of these 
monopoles can be described exclusively by the $SO(d)$ isovector Higgs fields in the {\em regular} gauge,
identifying the C-P charge with the Higgs field winding number. All this was carried out in all, even
and odd, dimensions.

Just as the magnetic Maxwell field, which is the reduced C-S form
pertaining to the monopole of the three dimensional GG model, can be identified 
as the Dirac monopole field strength, we have identified the reduced C-S forms, descended
from higher order C-P densities, as the Kalb--Ramond (KR) monopole fields of Nepomechie~\cite{N}
in all odd dimensions. The role analogous to that of the
't~Hooft--Polyakov monopole is played by the solitons of the gGG models, which we have 
called gGG monopoles. This construction is not possible in even dimensions.

Just as in the three dimensional case, the study of the gGG monopoles and the
corresping C-S forms, both in the {\em regular} and {\em Dirac} gauges, led
us to define the 't~Hooft tensors in all odd dimensions. These are identified with
the relevant KR field strengths of the gGG monopoles. All our
concrete constructions were carried out using the hedgehog (spherically symmetric) gGG monopoles
in three and five dimensions, but the results hold generally.

We now discuss briefly, connections of our results with physically relevant problems. 

The first is concerned with the construction of dilute gases of gGG monopoles.
KR theory is linear in the sense that the sum of two solutions is again a solution to the
KR equations \re{KR2} and \re{KR6}. This allows the construction of a {\em dilute gas} of KR monopoles
in analogy to the Coulomb gas in usual Maxwell theory. This Coulomb gas was used in Polyakov's 
work~\cite{Pol} to construct a dilute gas of 't~Hooft--Polyakov monopoles of the usual GG model in three 
dimensions, yielding a mechanism for confinement in the resulting QCD toy model. One can thus expect that
Polyakov's construction can be adopted to the gGG models in odd dimensions, but unfortunately
not in even dimensions --- and particularly not in the physically interesting case $d=4$, where the
significance of the gGG model $\CE^{(2,4)}$ has recently been pointed out \cite{KTZ}. Indeed, the
construction of a dilute gas from KR monopoles in arbitrary dimensions was carried out long ago in
the context of lattice field theory~\cite{S,O,Pe}.

Perhaps of most topical interest is the relevance of our results to the modern concept of branes.
In usual Maxwell theory, the elementary electrically charged objects are pointlike (zero dimensional),
and their time evolution is described by a {\em worldline}. The Maxwell potential describing a magnetic 
field, on the other hand, is a $1$--form which can be integrated along the worldline of the electric charge.
This line integral, multiplied by the electric charge coupling constant $e$, yields the interaction 
energy (or, in Minkowskian spacetime, action) between the electric charge and the Maxwell magnetic 
field. Generalising this construction to higher dimensions, the electrically charged objects which couple to
$d-2$ form KR potentials must have a $d-2$ dimensional worldvolume such that the interaction between
the electrically charged object and the KR field is described as the integral of the KR potential over this
world volume. This means that the electrically charged object itself is a $d-3$ brane. Nepomechie~\cite{N}
has shown that the existence of a KR monopoles with magnetic charge $Q_m$ leads to a quantisation condition
for the electric charge of the branes which couple to the KR monopole field in the way described above.
Therefore the gGG monopoles in $2M+1$ dimensions which we have discussed yield
the quantisation of the elementary electric charge of $2M-2$ branes in analogy to the 't~Hooft--Polyakov
monopole which forces the quantisation of elementary electric point charges (which are $0$--branes). 
KR fields, on the other hand, arise naturally in the context of string theories \cite{KR} which are the 
background of all modern brane physics, so that we expect strings and D-branes to be the correct context
in which our constructions may be of some significance.

Finally, we discuss a generalisation of the gGG monopoles used in this work.

The dynamics of the systems studied was described by the so-called gGG models, and in particular the 
solitons they support in odd spatial dimensions are the gGG monopoles alluded to above. 
These gGG monopoles are the classical solutions to the Euler--Lagrange equations in the temporal gauge 
$\rA_0 =0$. When the odd spatial dimensions are restricted to $d=4M-1$, the gGG monopoles are in fact 
solutions to the relevant first order Bogomol'nyi or self--duality equations. These were discussed in 
Ref.~\cite{OBT} and the analytic proof of existence given in Ref.~\cite{Y}. These self--dual gGG monopoles, 
which are generalisations of the BPS monopoles in three ($M=1$) dimensions, are entirely suited for 
generalisation to the corresponding `dyons' following the prescription used by Julia and Zee~\cite{JZ} in 
the $M=1$ case. This task can be performed systematically and will be reported elsewhere. The question 
here is, whether such `dyons' can be described as KR fields?

\section*{Acknowledgements}

We are grateful to P. Orland for having brought to our attention references \cite{S,O,Pe}.
F.Z. acknowledges a Higher Education Authority (HEA) Postdoctoral Fellowship. 

\begin{appendix}

\section{Kalb--Ramond theory and magnetic monopoles}
\label{appKR}

In this Appendix, we summarise the theory of static Kalb--Ramond (KR) fields and 
KR monopoles, following Nepomechie \cite{N}.

\subsection{Free KR theory}

Kalb--Ramond (KR) theories \cite{KR} generalise Maxwell electromagnetism as they 
deal with higher--rank antisymmetric tensor field strength instead of the 
rank two Mawxwell field strength tensor.
In mathematical terminology, KR field strengths are real valued $r$--forms in $d+1$ Minkowskian
spacetime dimensions. Static KR theory involving $r$--forms on $d$ dimensional space $\R^d$ then,
generalises Maxwell magnetostatics. 

Here, we consider the special case $r=d-1$ in $d$ spatial dimensions and 
simply refer to this special case of magnetic KR theories as the 
``KR theory in $d$ dimensions'' since it is this
type of magnetic KR theory which we consider in the main part of this work.   
It is defined in terms of the magnetic KR field strength $(d-1)$--form 
\begin{equation}
\tB=\frac{1}{(d-1)!}\tcB_{\mu_1\ldots\mu_{d-1}}\d x^{\mu_1}\wedge\ldots\wedge
\d x^{\mu_{d-1}} = \tdB_{\mu}(*\d x^{\mu}), \quad
\tcB_{\mu_1\ldots\mu_{d-1}}:=\epsilon_{\nu\mu_1\ldots\mu_{d-1}}\tdB_{\nu}
\end{equation}
which is closed on $\R^d$,
\begin{equation}
\d \tB = 0 \quad \Leftrightarrow \quad \epsilon_{\mu_1\ldots\mu_d}
\partial_{\mu_1}\tcB_{\mu_2\ldots\mu_d} = 0
\quad \Leftrightarrow\quad \partial_{\mu}\tdB_{\mu} = 0.
\label{KR2}
\end{equation}
Poincar\'e's lemma then allows to introduce the KR potential $(d-2)$--form 
\begin{eqnarray}
&& \tA =
\frac{1}{(d-2)!}\,\tcA_{\mu_1\ldots\mu_{d-2}}\d x^{\mu_1}\wedge\ldots\wedge
\d x^{\mu_{d-2}}= \frac12\tdA_{\mu_1\mu_2}
*(\d x^{\mu_1}\wedge \d x^{\mu_2}), \nonumber \\
&& \tcA_{\mu_1\ldots\mu_{d-2}}  = 
\frac12\,\epsilon_{\mu_1\ldots\mu_{d-2}\nu_1\nu_2}\tdA_{\nu_1\nu_2}
\end{eqnarray}
such that
\begin{equation}
\tB = \d\tA \qquad \Leftrightarrow \qquad \tcB_{\mu_1\ldots\mu_{d-1}}
= \partial_{[\mu_1}\tcA_{\mu_2\ldots\mu_{d-1}]} \qquad \Leftrightarrow \qquad
\tdB_{\mu} = \partial_{\nu}\tdA_{\mu\nu}.
\label{KR3}
\end{equation}
$\tA$ is well defined up to an exact $(d-2)$--form $\d\tG$,
\begin{eqnarray}
&& 
\tG = \frac{1}{(d-3)!}\,
\tcG_{\mu_1\ldots\mu_{d-3}}\d x^{\mu_1}\wedge\ldots\wedge
\d x^{\mu_{d-3}} = \frac{1}{3!}\tdG_{\mu_1\mu_2\mu_3}
*(\d x^{\mu_1}\wedge \d x^{\mu_2}\wedge \d x^{\mu_3}), 
\nonumber \\
&&
\tcG_{\mu_1\ldots\mu_{d-3}} = \frac{1}{3!}\,
\epsilon_{\mu_1\ldots\mu_{d-3}\nu_1\nu_2\nu_3}
\tdG_{\nu_1\nu_2\nu_3}
\end{eqnarray}
which means that $\tB$ is invariant under KR transformations
\begin{equation}
\tA \mapsto \tA + \d\tG \quad \Leftrightarrow \quad
\tcA_{\mu_1\ldots\mu_{d-2}}\mapsto \tcA_{\mu_1\ldots\mu_{d-2}}+
\partial_{[\mu_1}\tcG_{\mu_2\ldots\mu_{d-2}]}
\quad \Leftrightarrow \quad
\tdA_{\mu_1\mu_2}\mapsto \tdA_{\mu_1\mu_2}+\partial_{\nu}\tdG_{\nu\mu_1\mu_2}
\label{KR4}
\end{equation}
and a KR energy functional is supposed to preserve this 
invariance.

Free KR theory is given by the energy functional
\begin{equation}
\mbox{E}_{KR}^{(d)}[\tB]:= \frac12 \int_{\R^d}\tB\wedge * \tB := \frac{1}{2(d-1)!}
\int_{\R^d}\tcB_{\mu_1\ldots\mu_{d-1}}^2 = \frac12 \int_{\R^d}
\tdB_{\mu}^2.
\label{KR5}
\end{equation}
Varying $\mbox{E}_{KR}^{(d)}$ with respect to $\tA$ yields the magnetic 
KR equation
\begin{equation}
* \d\!*\!\tB = 0 \quad \Leftrightarrow \quad 
\partial_{\nu}\tcB_{\nu\mu_1\ldots\mu_{d-2}} = 0 \quad \Leftrightarrow \quad 
\partial_{[\mu_1}\tdB_{\mu_2]} = 0.
\label{KR6}
\end{equation}
Eqs.\ \re{KR2} and\re{KR6} are the $d$--dimensional generalisation of the
magnetic Maxwell equations in $d=3$ space dimensions.

\subsection{KR monopoles in $d$ dimensions}
\label{secKRmon}

Assuming the existence of `magnetic' charges changes the mathematical structure of KR theory. 
From the analog of Maxwell magnetostatics, the flux of the magnetic KR field strength form through the 
sphere $S^{d-1}$ equals a magnetic KR charge $Q_m$,
\begin{equation}
Q_m = \int_{S^{d-1}} \tB.
\label{KR7}
\end{equation}
Since $S^{d-1}$ is not simply connected, $\d\tB = 0$ does not enforce the existence of a 
global KR potential form. However, on the simply connected upper and lower
half spheres $S^{d-1}_+,S^{d-1}_-\subset S^{d-1}$,
KR potential forms exist, $\tB|_{S^{d-1}_{\pm}} = \d\dtA$. 
In order to yield a well--defined field strength form 
$\tB$, the potential forms ${}^+\!\tA$ and ${}^-\!\tA$ on the overlap of their definition ranges, i.e.\
the equator $S^{d-1}_+\cap S^{d-1}_- = S^{d-2}$, differ by a
{\em transition form} $\ttG={}^+\!\tA|_{S^{d-2}} - {}^-\!\tA|_{S^{d-2}}$
which has to be closed, $\d\ttG=0$.  The  magnetic KR charge is determined by $\ttG$,
\begin{equation}
Q_m = \int_{S^{d-1}}\tB = \int_{S^{d-1}_+}\d{}^+\!\tA + 
\int_{S^{d-1}_-}\d{}^-\!\tA  = \int_{S^{d-2}}\left({}^+\!\tA - 
{}^-\!\tA\right)=\int_{S^{d-2}} \ttG.
\label{KR12}
\end{equation}
In analogy to the corresponding construction for the Dirac monopole,
eq.\ (\ref{KR12}) can be called the Wu--Yang construction for magnetic KR charge.
The crucial point in the construction is that $\ttG$ is not exact on the equator $S^{d-2}$ 
although it is closed.

An actual solution $\mtB$ of the KR equations \re{KR2} and \re{KR6} on $S^{d-1}$ with nonvanishing flux
\re{KR7} is called a (magnetic) {\em KR monopole}. One can show \cite{N,EGH} that the only such solution
is given by
\begin{equation}
\mtB = \frac{Q_m}{\sphere_{d-1}}\frac{\hxm}{r^{d-1}}
(* \d x^{\mu}).
\label{KR13}
\end{equation}
with $\mtB = \d\dmtA$ on $S^{d-1}_{\pm}$, respectively. The explicit expressions for
$\dmtA$, which do not concern us here, are given in Ref.~\cite{N}.
Interpreting this as $(d-1)$--form on $\R^d$ in distributional sense (i.e.\
including the singularity at the origin) and using
\begin{equation}
\partial_{\mu}\left(\frac{\hx_{\mu}}{r^{d-1}}\right) = 
\sphere_{d-1}\delta(\x)
\label{KR15}
\end{equation}
yields
\begin{equation}
\d\mtB = Q_m \delta(\x) (*1),
\label{KR16}
\end{equation}
i.e.\ eq.\ (\ref{KR2}) is modified by a point charge acting as source 
for the KR field strength.

\section{Minimal generalised Georgi--Glashow models in $d=2,3,4,5$ dimensions}
\label{models}

In this Appendix we give the explicit expression for the energy densities, C-P densities and 
C-S densities/forms of the minimal gGG models in $d=2,3,4,5$ dimensions.

Using $\rS=-(\bPhi^2+\eins)$, the energy densities are
\begin{eqnarray}
\label{p=1,d=2}
\CE^{(1,2)} & = & \frac{1}{C_{(1,2)}}\tr \left[-\frac14\rF_{\mu\nu}^2 - \frac12 (\rD_{\mu}\bPhi)^2
+\frac12\rS^2 \right]
\\
\label{p=1,d=3}
\CE^{(1,3)} & = & \frac{1}{C_{(1,3)}}\tr\left[-\frac14\rF_{\mu\nu}^2 - \frac12 (\rD_{\mu}\bPhi)^2\right] 
\\
\CE^{(2,4)} & = & \frac{1}{C_{(2,4)}}\tr\left[\{\rF_{\mu[\nu}\rF_{\rho\sigma]}\}^2 + 
4\{\rF_{[\mu\nu},\rD_{\rho]}\bPhi\}^2
- 18  \left(\{\rS,\rF_{\mu\nu}\} + [\rD_{\mu}\bPhi,\rD_{\nu}\bPhi]\right)^2
- 54\{\rS,\rD_{\mu}\bPhi\}^2 + 54 \rS^4 \right] \nonumber \\
\label{p=2,d=4} \\
\label{p=2,d=5}
\CE^{(2,5)} & = & \frac{1}{C_{(2,5)}}\tr\left[\{\rF_{\mu[\nu}\rF_{\rho\sigma]}\}^2 + 
\{\rF_{[\mu\nu},\rD_{\rho]}\bPhi\}^2
-24 \left(\{\rS,\rF_{\mu\nu}\} + [\rD_{\mu}\bPhi,\rD_{\nu}\bPhi]\right)^2
-48 \{\rS,\rD_{\mu}\bPhi\}^2\right]\: .
\end{eqnarray}
$C_{(p,d)}>0$ are normalisation constants to be discussed below.

The corresponding residual C-P densities $\varrho^{(p,d)}\le\CE^{(p,d)}$ are found to be 
\begin{eqnarray}
\label{CP:p=1,d=2}
\varrho^{(1,2)} & = & \frac{1}{2C_{(1,2)}}i\epsilon_{\mu\nu}
\tr \left[\gamma_3\left(\rS\rF_{\mu\nu}+\rD_{\mu}\bPhi\rD_{\nu}\bPhi\right) \right]
\\
\label{CP:p=1,d=3}
\varrho^{(1,3)} & = & -\frac{1}{4C_{(1,3)}} \epsilon_{\mu\nu\rho}
\tr\left[\rF_{\mu\nu}\rD_{\rho}\bPhi \right] 
\\
\varrho^{(2,4)} & = & \frac{18}{C_{(2,4)}} \epsilon_{\mu\nu\rho\sigma}
\tr\left[\gamma_5\left(\rS^2\rF_{\mu\nu}\rF_{\rho\sigma}
-2\rS\{\rF_{\mu\nu},\rD_{\rho}\bPhi\}\rD_{\sigma}\bPhi
+2\left(\rS\rF_{\mu\nu} + \rD_{\mu}\bPhi\rD_{\nu}\bPhi\right)
\left(\rS\rF_{\rho\sigma} + \rD_{\rho}\bPhi\rD_{\sigma}\bPhi\right)
\right) \right] \nonumber \\
\label{CP:p=2,d=4}
\\
\label{CP:p=2,d=5}
\varrho^{(2,5)} & = & -\frac{24}{C_{(2,5)}} i \epsilon_{\mu\nu\rho\sigma\tau} 
\tr\left[3\rS\rF_{\mu\nu}\rF_{\rho\sigma}\rD_{\tau}\bPhi 
+2\rF_{\mu\nu}\rD_{\rho}\bPhi\rD_{\sigma}\bPhi\rD_{\tau}\bPhi\right].
\end{eqnarray}
The volume integral \re{gg2} of the C-P densities yields the C-P charge $q^{(p,d)}$. It is usual
to choose the normalisation constants $C_{(p,d)}$ such that the hedgehog \re{Ansatz} has unit topological
charge.

The C-P charges can be written as total divergences of the C-S densities $\Omega^{(p,d)}_{\lambda}$,
\begin{equation}
\varrho^{(p,d)} = \partial_{\lambda}\Omega^{(p,d)}_{\lambda}.
\end{equation}
The corresponding explicit expressions for the C-S densities are  
\begin{eqnarray}
\label{cCS:p=1,d=2}
\Omega^{(1,2)}_{\lambda} & = & -\frac{1}{C_{(1,2)}}i\epsilon_{\lambda\mu} 
\tr \left[\gamma_3 \left(\rA_{\mu} - \frac12\bPhi\rD_{\mu}\bPhi\right) \right]
\\
\label{cCS:p=1,d=3}
\Omega^{(1,3)}_{\lambda} & = &-\frac{1}{4C_{(1,3)}} \epsilon_{\lambda\mu\nu}
\tr\left[\bPhi\rF_{\mu\nu} \right] 
\\
\Omega^{(2,4)}_{\lambda} & = & -\frac{108}{C_{(2,4)}}\epsilon_{\lambda\mu\nu\rho}
\tr\left[\gamma_5\left(\rF_{\mu\nu}\rA_{\rho}-\frac23\rA_{\mu}\rA_{\nu}\rA_{\rho}
- \frac12 (2\eins+\bPhi^2)\bPhi\{\rF_{\mu\nu},\rD_{\rho}\bPhi\} 
+ \frac13\bPhi\rD_{\mu}\bPhi\rD_{\nu}\bPhi\rD_{\rho}\bPhi
\right)\right] \nonumber \\ 
\label{cCS:p=2,d=4}
\\
\label{cCS:p=2,d=5}
\Omega^{(2,5)}_{\lambda} & = & \frac{24}{C_{(2,5)}}i\epsilon_{\lambda\mu\nu\rho\sigma}
\tr\left[\left(3\eins+\bPhi^2\right)\bPhi\rF_{\mu\nu}\rF_{\rho\sigma}-
2\bPhi\rD_{\mu}\bPhi\rD_{\nu}\bPhi\rF_{\rho\sigma}\right].
\end{eqnarray}

Note that in the $d=2$ formulas \re{p=1,d=2},\re{CP:p=1,d=2} and \re{cCS:p=1,d=2}, the gauge field
$\rA_{\mu}=-iA_{\mu}\gamma_3$ is Abelian, and the Higgs field then has only two components as seen from
\re{higgs}, or alternatively can be parametrised by a single complex field $\varphi$ in \re{comps}. It is
then obvious that \re{p=1,d=2} pertains to the usual Abelian Higgs model.

Finally, we give the explicit expressions for the C-S forms $\bOmega^{(p,d)}$
of the four minimal gGG models \re{p=1,d=2}--\re{p=2,d=5} in the language of differential forms used 
in this work. They are defined as Hodge duals of the C-S density $1$--forms with components 
$\Omega^{(p,d)}_{\lambda}$ given by \re{cCS:p=1,d=2}--\re{CS:p=2,d=5},
\begin{equation}
\bOmega^{(p,d)}=\Omega^{(p,d)}_{\lambda}(*\d x^{\lambda}),
\end{equation}
thus
\begin{eqnarray}
\label{CS:p=1,d=2}
\bOmega^{(1,2)} & = & -\frac{1}{C_{(1,2)}}i
\tr \left[\gamma_3 \left(\bA - \frac12\bPhi\bD\bPhi\right) \right]
\\
\label{CS:p=1,d=3}
\bOmega^{(1,3)} & = &-\frac{1}{2C_{(1,3)}} 
\tr\left[\bPhi\bF \right] 
\\
\bOmega^{(2,4)} & = & -\frac{216}{C_{(2,4)}}
\tr\left[\gamma_5\left(\bF\wedge\bA-\frac13\bA\wedge\bA\wedge\bA 
- \frac12 (2\eins+\bPhi^2)\bPhi\{\bF,\bD\bPhi\}
+ \frac16\bPhi\bD\bPhi\wedge\bD\bPhi\wedge\bD\bPhi
\right)\right] \nonumber \\
\label{CS:p=2,d=4}
\\
\label{CS:p=2,d=5}
\bOmega^{(2,5)} & = & \frac{96}{C_{(2,5)}}i
\tr\left[\left(3\eins+\bPhi^2\right)\bPhi\bF\wedge\bF-
\bPhi\bD\bPhi\wedge\bD\bPhi\wedge\bF\right].
\end{eqnarray}

\end{appendix}


\bigskip
\bigskip








\bigskip
\bigskip


\begin{thebibliography}{99}

\bibitem{D}
P.A.M. Dirac, Proc.\ Roy.\ Soc.\ {\bf A 133} (1934) 60.

\bibitem{WY}
T.T. Wu and C.N. Yang, J. Math.\ Phys.\ {\bf 15} (1974) 53.

\bibitem{tH} 
G. 't~Hooft, Nucl.\ Phys.\ {\bf B 97} (1974) 276.

\bibitem{P}
A.M. Polyakov, JETP Lett.\ {\bf 20} (1974) 194.

\bibitem{AFG}
J.Arafune, P.G.O. Freund and C.J. Goebel, J. Math.\ Phys.\ {\bf 16} (1975) 433.

\bibitem{N}
R. Nepomechie, Phys.\ Rev.\ {\bf D 31} (1985) 1921.

\bibitem{KR}
M. Kalb and P. Ramond, Phys.\ Rev.\ {\bf D 9} (1974) 2273.

\bibitem{S}
R. Savit, Phys.\ Rev.\ Lett.\ {\bf 39} (1977) 55.

\bibitem{O}
P. Orland, Nucl.\ Phys.\ {\bf B 205} (1982) 107.

\bibitem{Pe}
R.B. Pearson, Phys.\ Rev.\ {\bf D 26} (1982) 2013.

\bibitem{Pol}
A.M. Polyakov, Phys.\ Lett.\ {\bf B 59} (1975) 82;
Nucl.\ Phys.\ {\bf B 120} (1978) 249.

\bibitem{T}
D.H. Tchrakian, J. Math.\ Phys.\ {\bf 21} (1980) 166;
Phys.\ Lett.\ {\bf B 150} (1985) 360; Int.\ J. Mod.\ Phys.\ A (Proc.\ Suppl.) {\bf 3A} (1993) 584.

\bibitem{T&OC}
G.M. O'Brien and D.H. Tchrakian, Mod.\ Phys.\ Lett.\ {\bf A 4} (1989) 1389; A. Chakrabarti and
D.H. Tchrakian, J.\ Math.\ Phys.\ {\bf 32} (1991) 2532.

\bibitem{T&OMS}
D. O'  S\'e, T.N. Sherry and D.H. Tchrakian, J. Math.\ Phys.\ {\bf 27} (1986) 325;
Zhong--Qi Ma, G.M. O'Brien and D.H. Tchrakian, Phys.\ Rev.\ {\bf D 33} (1986) 1177;
Zhong--Qi Ma, and D.H. Tchrakian, Phys.\ Rev.\ {\bf D 38} (1988) 3827; 
G.M. O'Brien and D.H. Tchrakian, Mod.\ Phys.\ Lett.\ {\bf A 4} (1989) 1389.

\bibitem{sph}
D.H. Tchrakian, Phys.\ Lett.\ {\bf B 150} (1985) 360.

\bibitem{ax}
A. Chakrabarti, T.N. Sherry and D.H. Tchrakian, Phys.\ Lett.\ {\bf B 162} (1985) 143;
J. Burzlaff, A. Chakrabarti and D.H. Tchrakian, J. Math.\ Phys.\ {\bf 34} (1993) 1665;
D.H. Tchrakian, Joel Spruck and Yisong Yang, Commun.\ Math.\ Phys.\ {\bf 188} (1997) 737.

\bibitem{KOT}
B. Kleihaus, D. O'Keeffe and D.H. Tchrakian, Phys.\ Lett.\ {\bf B 427} (1998) 327.

\bibitem{MT} Zhong--Qi Ma and D.H. Tchrakian, Lett.\ Math.\ Phys.\ {\bf 26} (1992) 179.

\bibitem{Zumino} B. Zumino, ``Chiral anomalies and differential geometry'',
{\bf in:} B.S. DeWitt and R.Stora (eds.), ``Relativity, groups and 
topology II'', Les Houches Session XL 1983, Elsevier 1984.

\bibitem{KTZ}
B. Kleihaus, D.H. Tchrakian and F. Zimmerschied, Phys.\ Lett.\ {\bf B 461} (1999) 224.

\bibitem{OBT}
D. O'S\' e, J. Burzlaff and D.H. Tchrakian, Lett.\ Math.\ Phys.\ {\bf 13} (1987) 211.

\bibitem{Y}
Yisong Yang, Lett. Math. Phys. {\bf 19} (1990) 257; {\it ibid.} {\bf 20} (1990) 285.

\bibitem{JZ}
B. Julia and A. Zee, Phys.\ Rev.\ {\bf D 11} (1975) 2227.

\bibitem{EGH}
T. Eguchi, P.B. Gilkey and A.J. Hanson, Phys.\ Rep.\ {\bf 66} (1980) 213. 





\end{thebibliography}
\end{document}